%% file: Final.tex
 \documentclass[journal, comsoc]{IEEEtran}

\usepackage{gensymb}
\usepackage{cite}
\usepackage{amsmath,amssymb,amsfonts}
\usepackage{graphicx}
\usepackage{xcolor}
\usepackage{kotex}  
\usepackage{caption}
\usepackage{subcaption}
\usepackage{caption}
\usepackage{subcaption}
\usepackage[font=scriptsize]{caption}
\usepackage{stfloats}
\usepackage{float}
\usepackage{wrapfig}
\usepackage{xcolor}
\usepackage{colortbl} 
\usepackage{color,soul}
\usepackage{url}
\usepackage{verbatim}
\usepackage{graphicx}
\usepackage{cite}

\usepackage{multirow}
\usepackage{array,boldline, makecell,booktabs}
\usepackage{longtable} 
\usepackage{acronym}

    \usepackage[group-separator={,},group-minimum-digits=4]{siunitx}

\begin{document}
%
\title{An Overview of Intelligent Meta-surfaces for 6G and Beyond: Opportunities, Trends, and Challenges}

%
%
%

\author{Mayur Katwe, Aryan Kaushik, Lina Mohjazi, Mohammad Abualhayja'a, Davide Dardari, Keshav Singh, Muhammad Ali Imran, M. Majid Butt, and Octavia A. Dobre
\thanks{M. V. Katwe is with the National Institute of Technology, Raipur, India (e-mail: mvkatwe.ece@nitrr.ac.in).\\
$~~~$A. Kaushik is with the School of Engineering \& Informatics, University	of Sussex, UK (e-mail: aryan.kaushik@sussex.ac.uk). 
\\
$~~~$L. Mohjazi, M. Abualhayja'a, and M. A. Imran are with the James Watt School of Engineering, University	of Glasgow, UK (e-mails: \{lina.mohjazi, muhammad.imran\}@glasgow.ac.uk, m.abualhayjaa.1@research.gla.ac.uk). \\
$~~~$D. Dardari is with the 
    Dipartimento di Ingegneria dell'Energia Elettrica e dell'Informazione ``Guglielmo Marconi" (DEI), 
   University of Bologna, and WiLab-CNIT,  
   Bologna (BO), Italy (e-mail: davide.dardari@unibo.it).
\\
$~~~$K. Singh is with the Institute of communications Engineering, National Sun Yat-sen University, Taiwan (e-mail: keshav.singh@mail.nsysu.edu.tw). \\
$~~~$M. M. Butt is with Nokia, Naperville, IL, USA. (e-mail: majid.butt@nokia.com).\\
$~~~$O. A. Dobre is with the Faculty of Engineering and Applied Science, Memorial University, Canada (e-mail: odobre@mun.ca).
}
}

%
%

\markboth{Submitted to IEEE Communications Standards Magazine}%
{Shell \MakeLowercase{\textit{et al.}}: Bare Demo of IEEEtran.cls for IEEE Journals}

\input{Acronym}

\maketitle

\begin{abstract}
With the impending arrival of the sixth generation (6G) of wireless communication technology, the telecommunications landscape is poised for another revolutionary transformation. At the forefront of this evolution are intelligent meta-surfaces (IS), emerging as a disruptive physical layer technology with the potential to redefine the capabilities and performance metrics of future wireless networks. As 6G evolves from concept to reality, industry stakeholders, standards organizations, and regulatory bodies are collaborating to define the specifications, protocols, and interoperability standards governing IS deployment. Against this background, this article delves into the ongoing standardization efforts, emerging trends, potential opportunities, and prevailing challenges surrounding the integration of IS into the framework of 6G and beyond networks. Specifically, it provides a tutorial-style overview of recent advancements in IS and explores their potential applications within future networks beyond 6G. Additionally, the article identifies key challenges in the design and implementation of various types of intelligent surfaces, along with considerations for their practical standardization. Finally, it highlights potential future prospects in this evolving field.

\end{abstract}


\IEEEpeerreviewmaketitle

\vspace{-3mm}

\section{Introduction }
\label{sec:1}
From the early days of second-generation (2G) to the current era of fifth-generation (5G) deployments, each generational leap has brought  significant advancements in telecommunication architecture, spectrum utilization, data rates, latency, reliability, and user connectivity. As we approach the era of sixth-generation (6G), the focus shifts towards harnessing emerging technologies that can unlock new dimensions of connectivity and efficiency. In line with the core concept of 6G's vision for ubiquitous connectivity,  intelligent meta-surfaces \acp{IS}, also known as reconfigurable intelligent surfaces (RIS), represent a paradigm shift in wireless communication by endowing passive surfaces with active control over signal propagation\cite{wen2024shaping}. 
Their capabilities are extensive, including reflections, refractions, absorptions,  beam splitting, polarization manipulation, and focusing, collectively reshaping and directing electromagnetic waves in the radio propagation environment \cite{liu2022a}. 
Beyond these fundamental manipulations, IS facilitates analog processing of \ac{EM} waves, offering a unique low-complexity, low energy consumption, and low-latency approach to signal modification and enhancement and holds immense promise across a myriad of application domains, ranging from ultra-dense urban environments to remote rural areas and beyond \cite{wu2020towards}. 
\begin{figure}[t!]
\centerline{\includegraphics[width=\columnwidth]{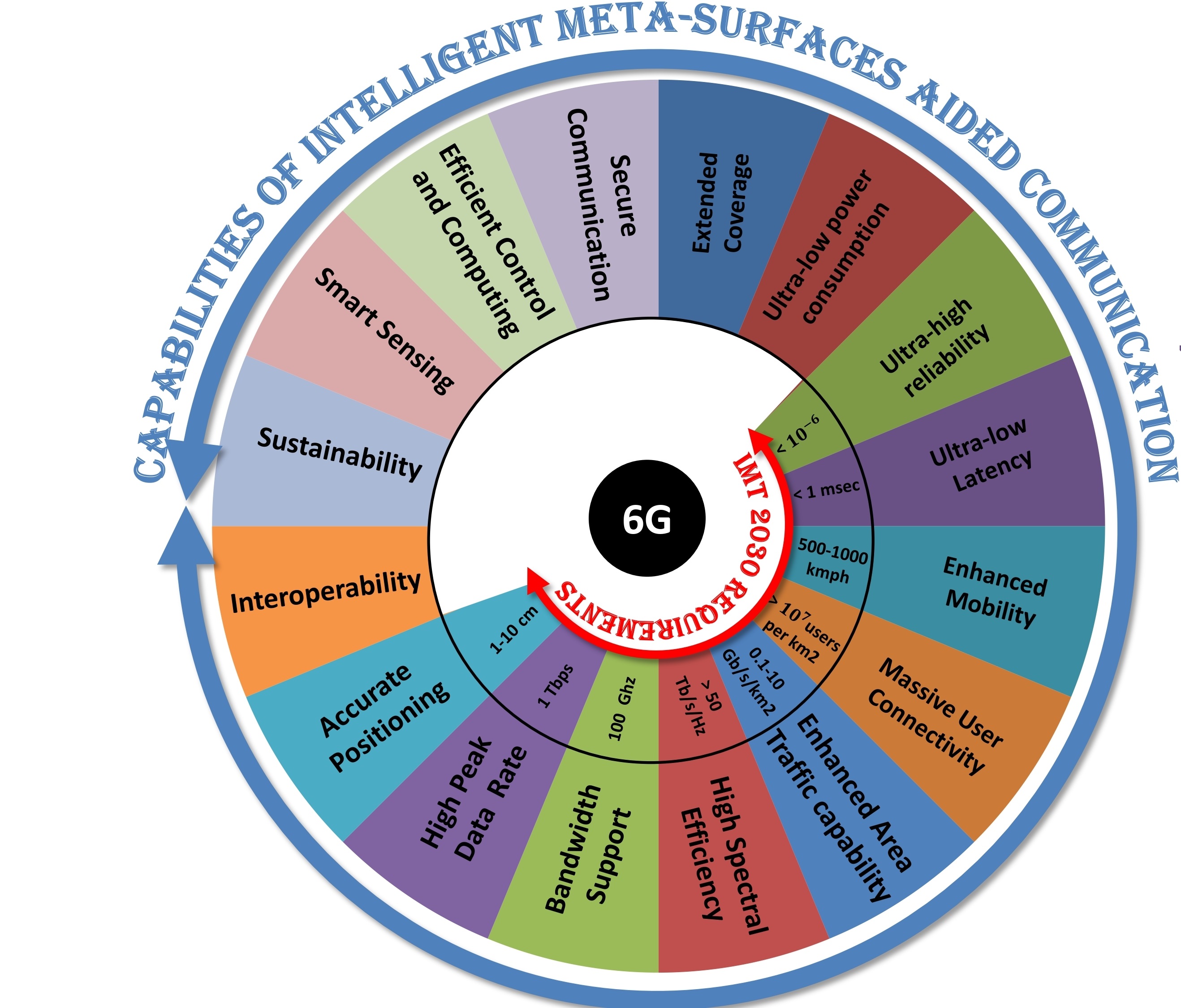}}
	\caption{IMT-2030 Requirement and IS Capabilities.}
	\label{fig1_imt}
	\vspace{-0.55em}
\end{figure}

IS-aided communication holds immense potential in fulfilling the ambitious requirements set forth by the International Mobile Telecommunications-2030 (IMT-2030) for 6G networks, as illustrated in Fig. \ref{fig1_imt}. In particular, 6G networks aim to serve denser user populations, deliver higher data rates, and ensure ultra-low latency with exceptional reliability. With IS technology, these goals become achievable, allowing networks to substantially enhance their capabilities, possibly in a sustainable way. For instance, 6G networks are projected to serve higher numbers of users, reaching up to $10^7$ users per square kilometer, while delivering data rates of 1 Gbps in the downlink. Achieving an end-to-end communication latency of less than 1 milliseconds with 99.99999$\%$ network reliability is also a crucial target.

\subsection{IS Empowering Multifunctions: Control, Computing, Communication, Localization, Sensing}
Undoubtedly, IS serves as a game-changer by dynamically manipulating electromagnetic waves to optimize signal propagation, extend coverage, and mitigate interference, thereby enhancing connectivity and reliability for a diverse range of services and users.
\begin{enumerate}
\item \textbf{Enhanced Mobile Broadband and Reliability and low latency communication (mBRLLC)}: In 3rd Generation Partnership Project (3GPP) Release (Rel.) 15, two core services, namely enhanced mobile broadband (eMBB) and ulra-relaible low latency communication (URLLC), were introduced, with ongoing refinement continuing through Rel. 19. While eMBB prioritizes ultra-high data rates, URLLC focuses on achieving low latency and ultra-high reliability. As wireless applications evolve, there is a growing need for extended capabilities, called Enhanced Mobile Broadband and Reliability and low latency communication (mBRLLC) \cite{saad2020vision},  which can support support high data rates as well as both ultra-high reliability, and minimal end-to-end latency. Owing to their reconfigurability and extra degree of freedom in terms of beamforming, IS-supported communication effectively meets these demanding requirements, particularly in resource-limited environments.
\item \textbf{Sensing and Localization}: Beyond their conventional roles, IS can be also empowered to perform integrated functions, such as data transmission, sensing, localization/tracking, imaging, and detection of passive objects, even in non-line-of-sight (NLOS) channel conditions.  In particular, IS enhances localization and sensing capabilities by providing fine-grained control over signal reflections and propagation paths, enabling high-precision positioning and environmental monitoring essential for emerging applications such as augmented reality, autonomous vehicles, and smart infrastructure \cite{chepuri2023ISAC}.
\item \textbf{Control}: IS empowers precise and responsive management of network resources, enabling real-time adjustments and optimizations to adapt to changing environmental conditions and user requirements \cite{singh2023indexed}. Such unique application of IS is demonstrated by modulating self-conjugating meta-surfaces (MSCM), designed to conjugate the complex envelope of incoming signals and embed information, enabling automatic alignment with the direction of an incoming signal, even without requiring channel state information (CSI) estimation \cite{DarLotDecPas:J23}.
\item \textbf{Multi-Access Computing (MEC)}: The effective deployment of multi-access edge computing for private networks and industrial networks, as discussed in several releases (Rel. 15, 17, 18), remains a challenge due to dynamic channel conditions.
In the realm of computing, IS can enhance edge computing by optimizing signal transmission and reception, reducing latency, and improving overall system efficiency. Additionally, IS can be integrated into distributed computing systems, leverage its ability to adjust signal characteristics to enhance communication. 
\item \textbf{Backward Compatibility and Interoperability}: 
Introduction of millimeter wave (mmWave) for cellular networks began with 3GPP Rel. 17 NR. The next phase of advanced 5G, i.e., Rel. 19 will feature massive multiple input multiple output (MIMO), although its effectiveness is limited at higher frequencies due to obstructions. IS technology addresses this by extending coverage, especially for high frequencies, without traditional relays. It seamlessly integrates with existing 5G infrastructure for smooth transitions to higher frequency bands, enhancing spectrum utilization and supporting diverse applications across different frequency ranges.


\end{enumerate}  

\section{Recent Advancements in IS}
\begin{figure*}[t!]
 \hfill
  \begin{subfigure}{.245\linewidth}
 \centering
		\includegraphics[height=3.4cm]{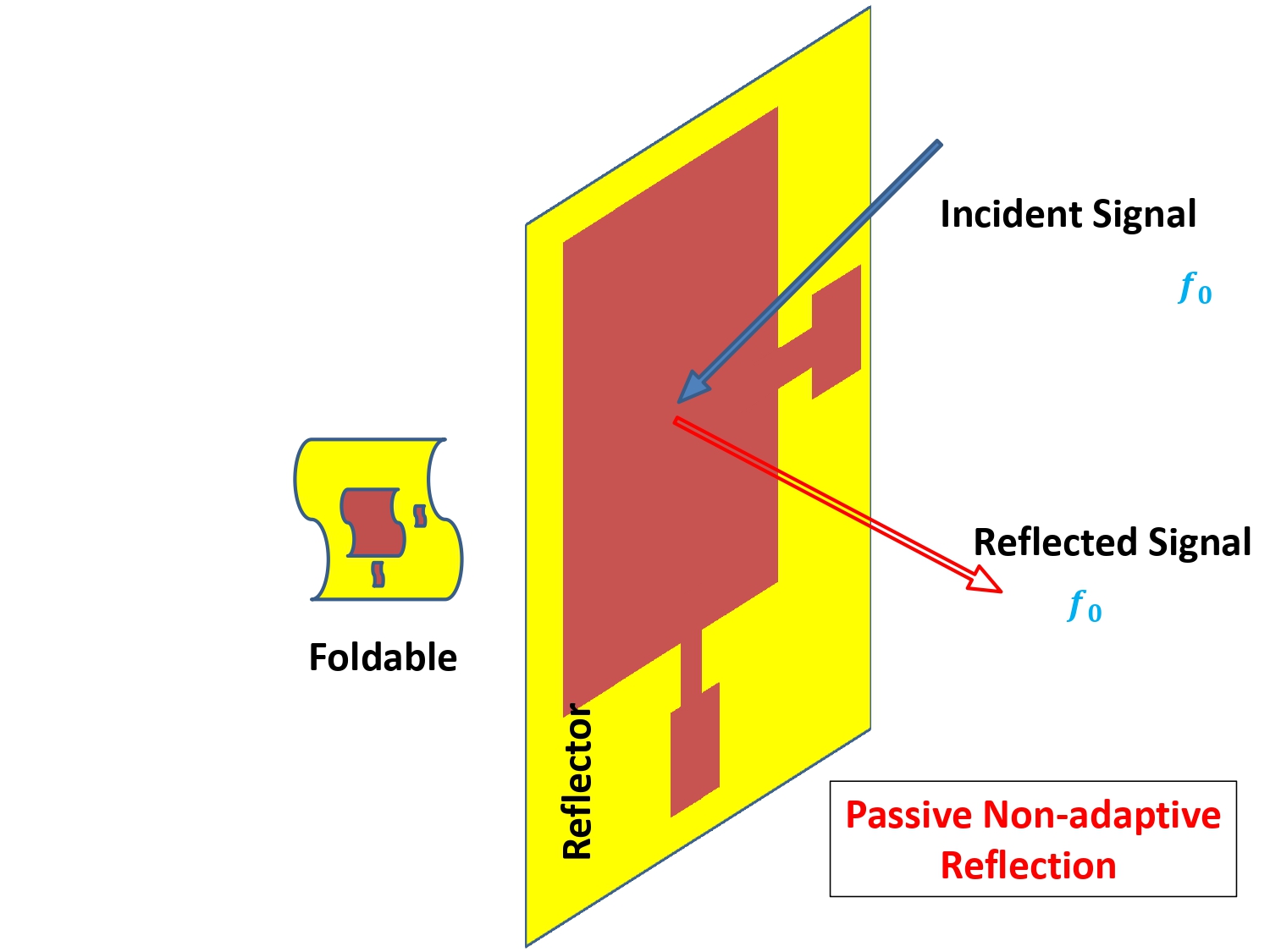}
		\caption{Non-Reconfigurable Surface}
		\label{2a}
	\end{subfigure}
 \hfill
 \begin{subfigure}{.245\linewidth}
 \centering
		\includegraphics[height=3.4cm]{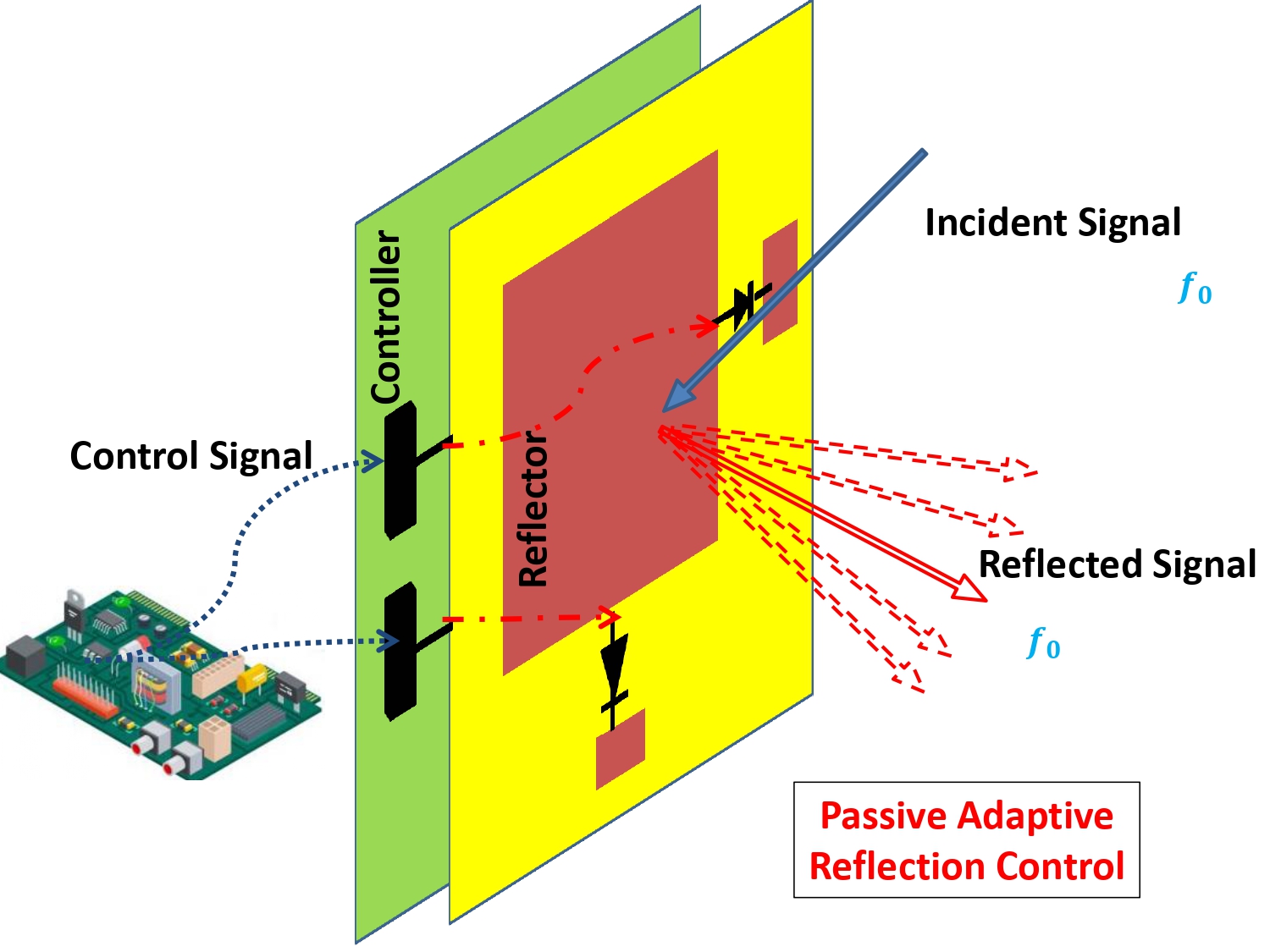}
		\caption{Passive RIS}
		\label{2b}
	\end{subfigure}
  \begin{subfigure}{.245\linewidth}
 \centering
		\includegraphics[height=3.4cm]{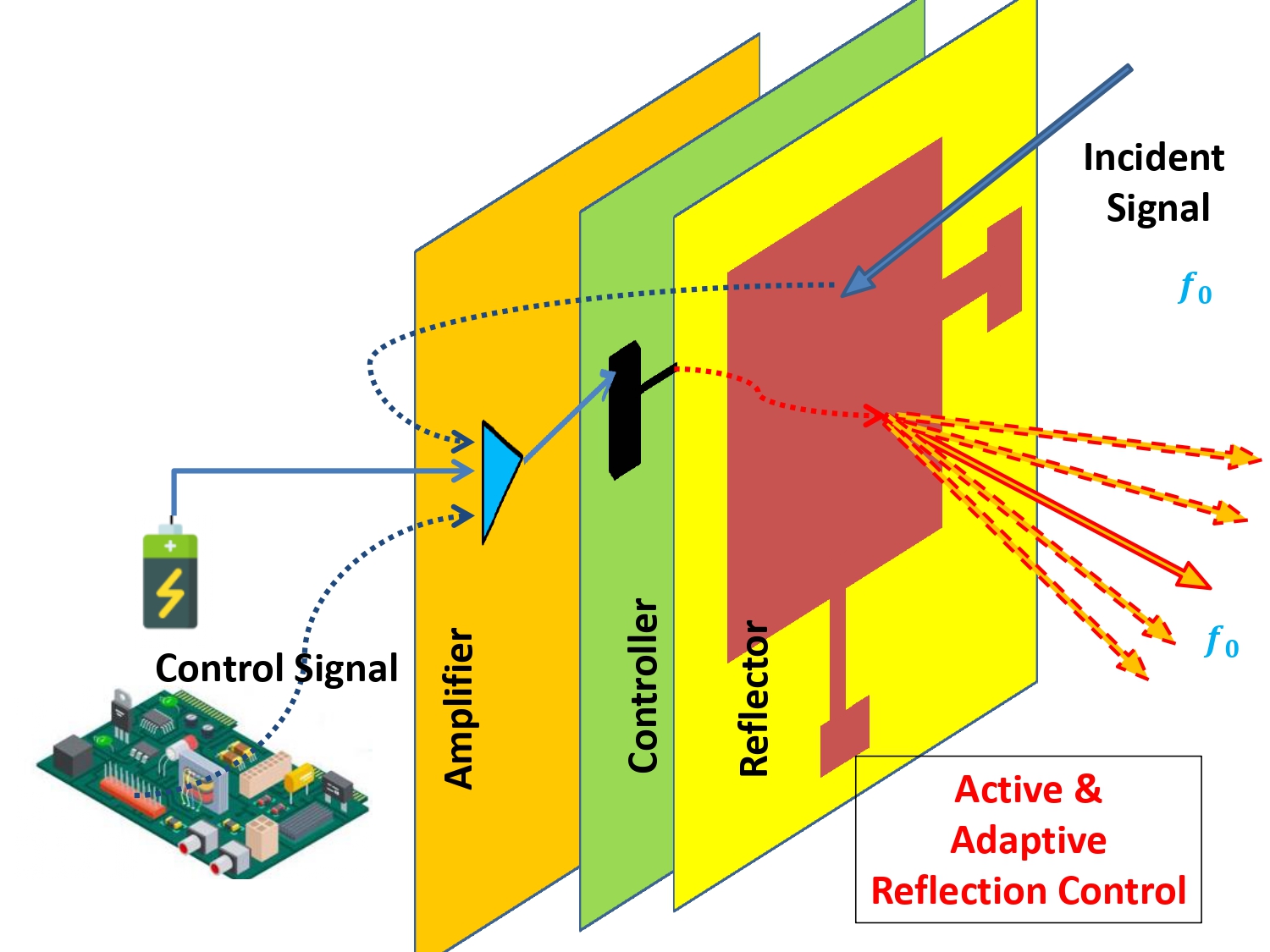}
		\caption{STAR-RIS}
		\label{2c}
	\end{subfigure}
  \begin{subfigure}{.245\linewidth}
 \centering
		\includegraphics[height=3.4cm]{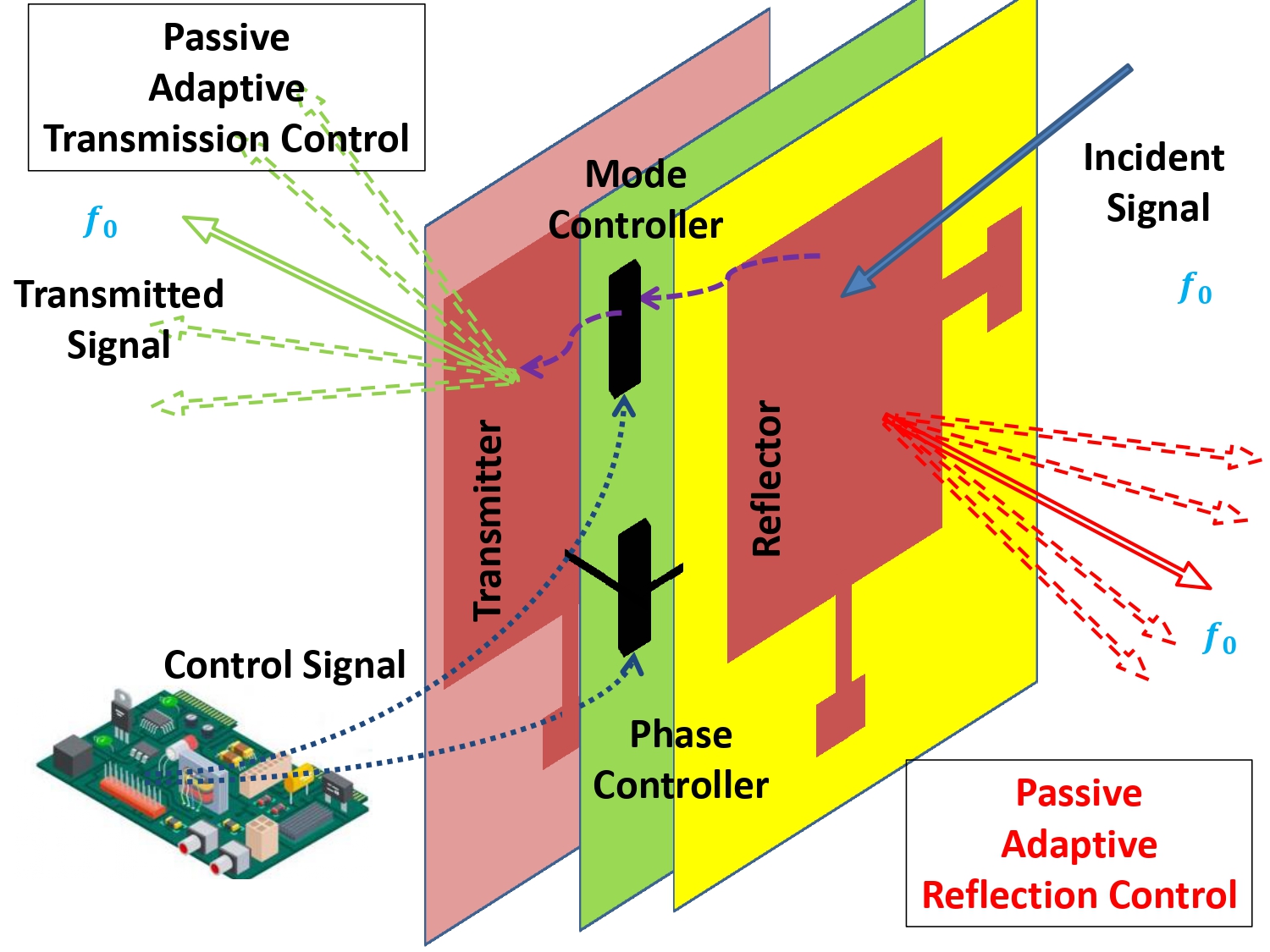}
		\caption{Active RIS}
		\label{2d}
	\end{subfigure}
  \begin{subfigure}{.35\linewidth}
 \centering
		\includegraphics[height=3.4cm]{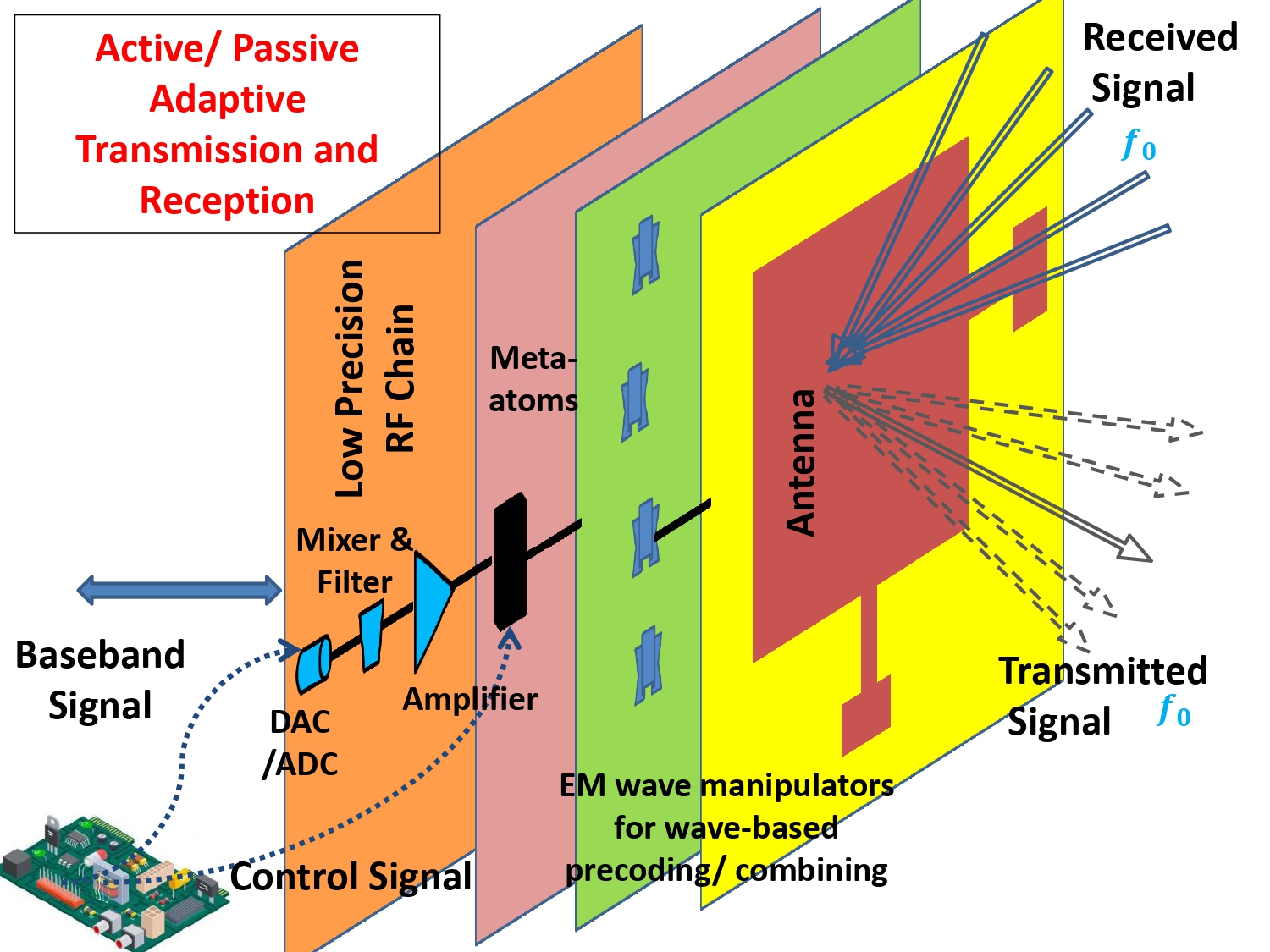}
		\caption{Stacked Intelligent Meta-surfaces (SIM)}
		\label{2e}
	\end{subfigure}
  \begin{subfigure}{.245\linewidth}
 \centering
		\includegraphics[height=3.4cm]{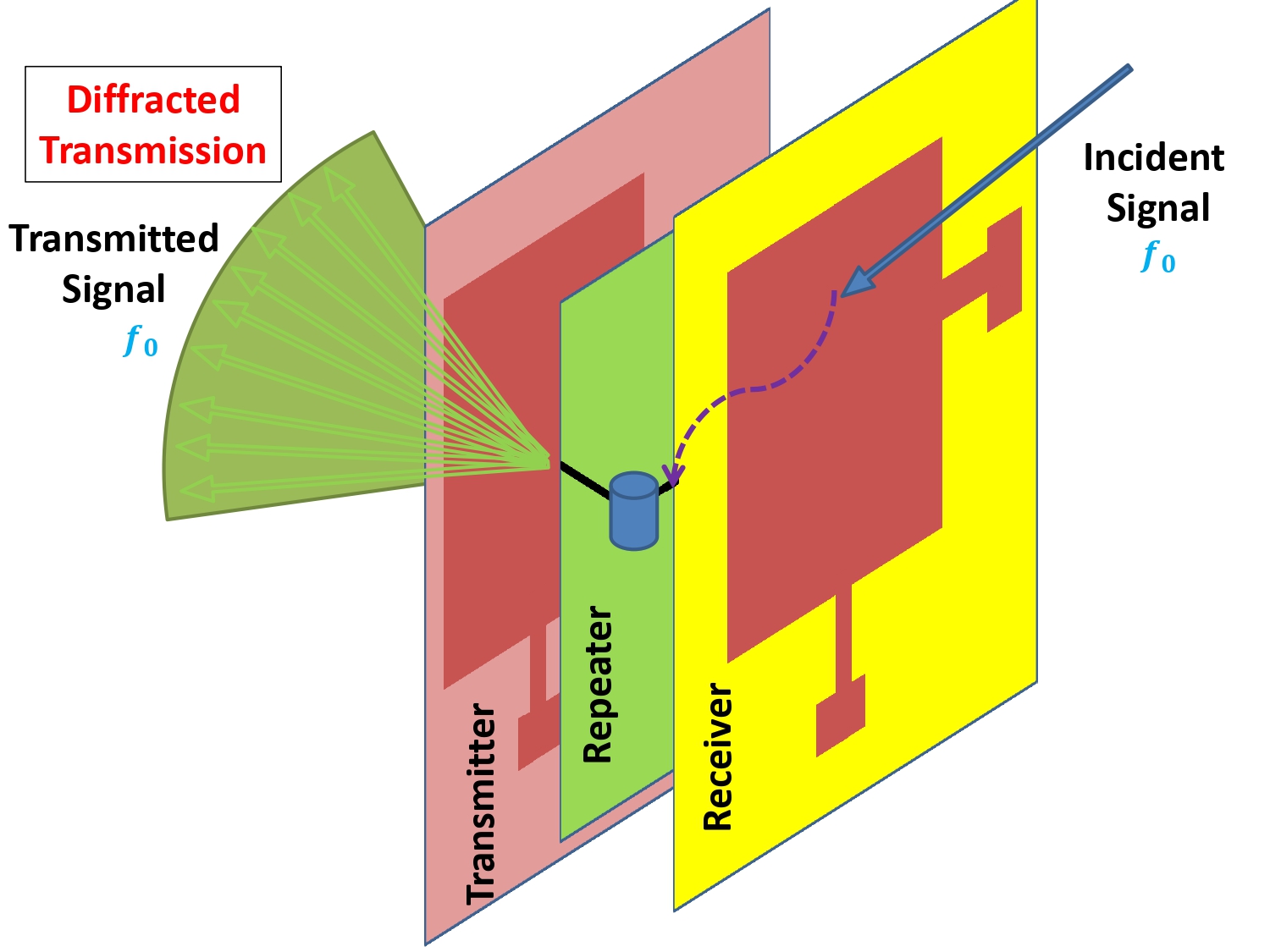}
		\caption{Transparent Meta-lens}
		\label{2f}
	\end{subfigure}
  \begin{subfigure}{.35\linewidth}
 \centering
		\includegraphics[height=3.4cm]{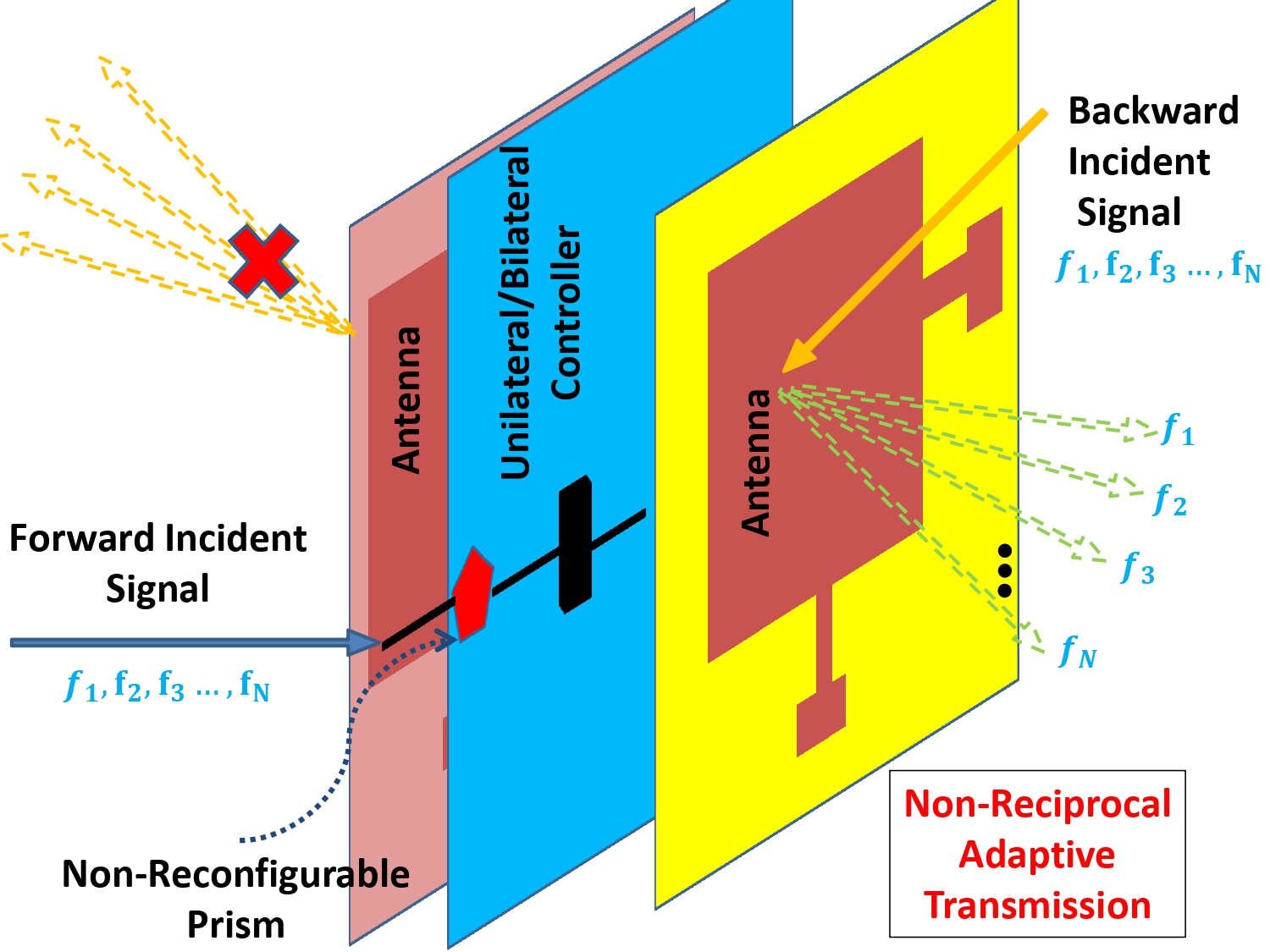}
		\caption{Meta-prism}
		\label{2g}
	\end{subfigure}
 \caption{Unit Cell of Various IS.}
\end{figure*}
 IS, essentially an array of small meta-material unit cells, can control incident, reflected, and transmitted electromagnetic fields. 
 In the last decade, significant advancements have been made in meta-materials, offering a diverse range of functionalities such as bandpass and bandstop frequency-selective surfaces, high-impedance surfaces, narrowband perfect absorbers, twist polarizers, right-handed circular-polarization frequency-selective surfaces, linear-to-circular polarization converters, two-dimensional leaky-wave antennas with conical-beam patterns, focusing transmit arrays, focusing reflect arrays, Luneburg lenses, holograms, and more \cite{li2022empowering}.
These functionalities have propelled the evolution of diverse IS unit cells, each possessing unique operational capabilities and underlying technologies, all aimed at enhancing 6G networks, as elucidated below:
\begin{enumerate}
    \item \textbf{Smart Skin}: 
A sophisticated "smart skin," meticulously designed without reliance on a dedicated power source and control mechanism,  embodies  the basic level of innovation for passive IS-aided communication. This ground-breaking concept features a non-reconfigurable reflective surface, elegantly depicted in Fig \ref{2a}. Typically fashioned into uniform tiles, these surfaces streamline production processes and allow for effortless mounting on structures of various shapes and sizes. While this adherence to uniformity simplifies manufacturing and operational aspects, it may pose challenges w.r.t. the standardized design, security concern, and their seamless integration into technically sophisticated environments.
\item \textbf{Dynamic IS antennas (or DMAs):} Another type of dynamic passive IS configurations, commonly referred to as DMAs, allows for configuring the phase shift of incoming signals on the reflector of each unit cell using a specialized control plane.  Each unit element's reflector may include single or multiple control elements like RF switches,PIN diodes, etc. Using either a micro-controller or field-programmable gate array, the controller plane sends analog or digital signals to these elements, altering the phase shift of incoming signals on the reflector to facilitate signal steering. For instance, N-bit discrete reconfigurability allows for phase shift variation with a resolution of $\frac{2^{n}}{2\pi}$.  Integrating DMAs presents  into base stations and access points can also enable highly adaptable signal processing.  

    \item \textbf{STAR-RIS}: Reflecting only IS render  limiting service to devices within the reflection space, i.e.,  only $180^\circ$ space. On the other hand, the concept of simultaneous transmitting and reflecting RIS (STAR-RIS) has emerged \cite{liu2021star}. This innovation allows for full-space (360$^\circ$) smart radio propagation by enabling both transmission and reflection of incident signals with beam steering, signal enhancement, or interference mitigation for full space, as illustrated in Fig. \ref{2c}.  Consequently, STAR-RIS exhibits enhanced flexibility and versatility in signal transmission compared to conventional RISs.
\item \textbf{Active RIS}:   Active RISs, also known as "smart repeaters," can amplify reflected signals using integrated amplifiers without any RF chains, thus mitigating the double path loss signal attenuation observed in passive IS \cite{zhang2023active}, as depicted in Fig. \ref{2d}.  Although requiring less power than conventional relays, the unit cell of active RISs necessitates additional power for amplifier operation, resulting in an extra power budget compared to passive RISs. Notably, active RIS configurations comes with varied amplification options, with fully connected setups allowing for individual power supplies and sub-connected setups ensuring uniform amplification.
    \item  \textbf{SIM }: Another variant of IS, stacked intelligent meta-surfaces (SIM), leverages principles from holography to create virtual multi-antenna arrays that can dynamically adapt to changing communication conditions. In other words,  a sophisticated combination of multiple stacked metasurface layers is used to fabricate a SIM, potentially surpassing the performance of single-layer metasurface counterparts like passive IS and metasurface lenses. In SIM, multiple RIS are stacked in order to perform more complex processing operations (eg. precoding and combining) directly in the EM domain  thus limiting the number of RF chains and hence the complexity of the antenna array typically deployed at the BS, as depicted in Fig. \ref{2e}.   
    \item  \textbf{Meta-lens}:  Recently, NTT DOCOMO experimented  a "transparent metasurface lens" to address challenges in  base station antenna coverage for indoors for better signal strength. They propose using  to concentrate weak radio waves passing through glass windows onto specific indoor locations, termed "focal points". This lens, attached to the indoor side of a glass window, amplifies signal power by focusing radio waves onto these focal points and providing "diffracted transmission", as shown in Fig. \ref{2e}. By installing repeaters, reflectors, or RIS at these points, coverage can be extended indoors. 
    \item \textbf{Meta-prism}:  A a reciprocal/nonreciprocal meta-surface-based prism,  inspired from the theory of optical prisms, refers to  comprised of an array of phase- and amplitude-gradient frequency-dependent spatially variant radiating multiple frequency EM signals, as shown in Fig. \ref{2f} for unilateral transmission. It is a metasurface prism which incorporates frequency-dependent spatially variant cells and that can be successfully exploited to provide some degrees of freedom in signal transmission/reflection without the need for reconfigurability and power supply \cite{DarMas:J21}.
    \item   \textbf{Beyond Diagonal RIS, RIS 2.0}:  Traditional  RIS is characterized by a diagonal phase shift matrix, whereas non-diagonal RIS, recently studied and referred to as RIS 2.0, introduce inter-element connections, adding circuit complexity and allowing for mathematical models beyond diagonal matrices \cite{li2024bdris}. 
    Diagonal matrices imply independent adjustments to each unit cell and local boundary conditions, whereas non-diagonal matrices denote interdependence among the phase shifts and global boundary conditions and allow for spatial selectivity. Spatial selectivity typically refers to the capability of reflecting a signal coming only from a specific direction, thus avoiding other signal sources generating reflections in unwanted directions.
\end{enumerate}

\begin{figure*}[t!]
\centerline{\includegraphics[width=\linewidth]{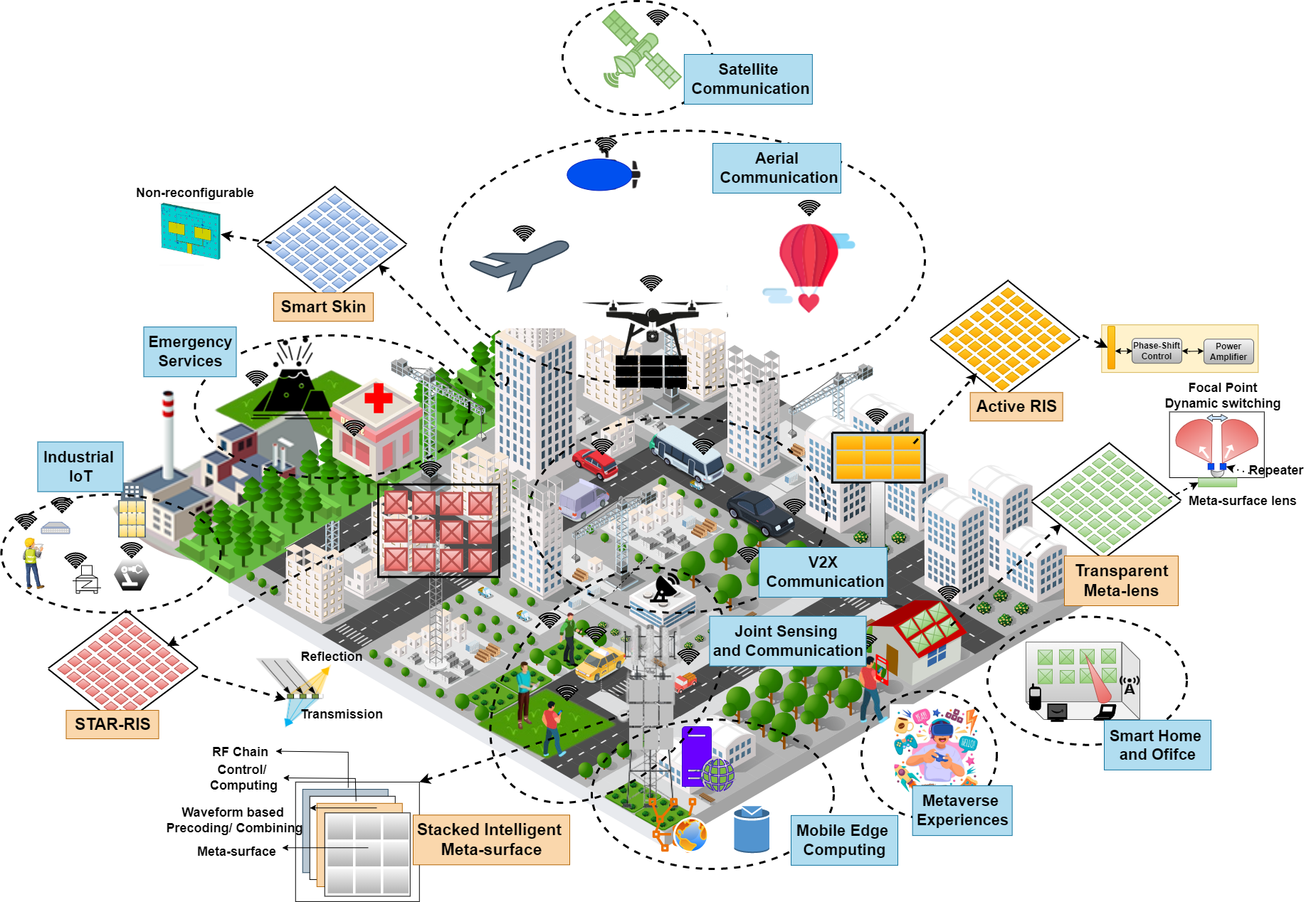}}
	\caption{Key Usage Scenarios of IS for 6G and Beyond.}
	\label{fig1}
\end{figure*}

\section{ Key Usage Scenarios and Applications of IS for 6G and Beyond}
\label{sec:4}
Thanks to advancements in IS, IS-assisted communication is poised to revolutionize the landscape of 6G and beyond applications as shown in Fig. \ref{fig1}, offering unparalleled opportunities for innovation and connectivity in the digital age. This sets the stage for the deployment of highly interconnected mobile, networks, boasting high data rates, ultra-high reliability, and ultra-low latency for various key usage scenarios for 6G and beyond as discussed below: 
\subsection{Integrated Sensing and Communications (ISAC)}
IS technology plays a pivotal role in realizing the potential of ISAC, particularly in enhancing communication systems and RF localization capabilities.  
The current design of IS aided ISAC systems  typically fall into two categories: waveform design strategies conducted independently of IS, and sensing-driven waveform design approaches that leverage IS for fine-tuning communication waveforms post-initial sensing operations. Achieving this entails striving to approximate the Pareto frontier of key sensing and communication metrics, including target location accuracy and signal-to-noise ratio (SNR) improvements for cellular users. 
Moreover, sophisticated hardware capabilities such as hybrid RIS and SIM systems exhibit significant potential for ISAC applications, providing an additional layer of sensing and actuation capabilities poised to elevate system performance.

\subsection{Non-Terrestrial Networks (NTN)}
In the realm of non-terrestrial networks (NTNs),  IS emerges as a transformative force, revolutionizing inter-communication among NTN nodes comprising UAVs, high altitude platform stations (HAPS), and satellites, particularly low earth orbit (LEO) satellites \cite{jamshed2024synergizing}. Here, UAVs, equipped with RISs (aerial RIS), can extend coverage and relay signals between terrestrial nodes and HAPS stationed above. HAPS, acting as hubs, receive signals from UAVs and terrestrial nodes, relaying them further. IS integration on HAPS mitigates signal attenuation and enhances transmission reliability. 
RISs offer a solution, enabling energy-efficient transmission and enhancing link reliability. In the coexistence of  HAPS, LEO and geostationary earth orbit (GEO) satellites, RISs can improve ground-to-air links and optimizing air-to-ground links, thus, boosting received signal strengths and transmit power efficiency. 

\subsection{Vehicle to Everything (V2X)}
Further, IS offers a transformative solution to enhance the performance and reliability of various V2X infrastructures  including vehicle-to-vehicle (V2V) communication for collision avoidance, vehicle-to-infrastructure (V2I) communication for traffic management and optimization, and vehicle-to-pedestrian (V2P) communication for enhanced pedestrian safety.
With IS deployed strategically, dynamic beamforming techniques can be employed to steer V2X signals toward their intended recipients while minimizing interference and maximizing signal strength. 
  Specifically, 
active RIS and STAR-RIS configurations further augment RIS-aided V2X communication.   While, STAR-RIS provides full 360-degree coverage, ensuring seamless communication in dynamic environments.  While IS integration hold potential for V2X, sophisticated channel-tracking and beamforming schemes are critical which can be resolved using AI-driven solutions such as data-driven deep learning techniques . 

\subsection{MEC for IoT devices}
Deploying IS alongside MEC heralds a transformative approach to optimizing network performance, especially for IoT devices. Placing IS units strategically at the network periphery, closer to internet of things (IoT) devices, streamlines data processing by reducing the distance data must travel, thereby significantly lowering latency—a crucial requirement for real-time IoT applications. Moreover, RIS-enhanced MEC improves reliability by establishing dependable connections between IoT devices and edge servers, addressing issues like packet loss and signal interference, even in challenging environments like urban or industrial settings \cite{bai2020latency}. This fosters efficient data aggregation, processing, and dissemination across fog nodes, ultimately enhancing scalability and performance for cloud-based applications and services such as data analytics, machine learning, and virtualization.  Importantly, adopting a low cost smart skin solution (as discussed earlier), offers a stable and cost-effective solution, ideal for IoT deployments in urban environments. 
\subsection{Metaverse Experiences}   Within this immersive realm, RIS deployment holds tremendous potential in revolutionizing the metaverse experience, offering immersive and seamless connectivity for virtual and augmented reality (AR/VR) applications. With RIS aiding metaverse experiences, the boundaries between physical and digital realities blur, opening up new possibilities for collaborative gaming, virtual events, remote work, and social interactions in the digital realm.  By leveraging AI and big data analytics, digital twins (DT), offering digital replicas of real-world products and systems, can process real-time data to create comprehensive models, enabling real-time decision-making and improving efficiency across various processes, including wireless communication environments empowered by RIS. This integration offers insights into dynamic communication environments, enhancing network reliability, reducing latency, and maximizing overall efficiency, thereby shaping the future of immersive metaverse experiences.
\subsection{Other Applications}
Beyond its primary applications, RIS technology introduces new possibilities for energy-efficient communication and networking opening possible avenues for wireless power transfer, cell-free massive MIMO (CF-mMIMO) systems operating at higher frequencies. Recent literature showcases that RIS can also empower semantic communication, enabling devices to exchange information not only based on traditional communication protocols but also by leveraging contextual knowledge and semantic understanding.  This innovative approach enhances communication efficiency, enabling more meaningful interactions between devices and enabling advanced applications such as context-aware services and intelligent decision-making. 

\section{Examples of IS-aided Use-cases}
This section describes case studies of IS-aided communication and offers insights into the diverse applications and prospects of IS in revolutionizing wireless communication.
\subsection{Aerial RIS-assisted HAP-terrestrial communication}
Firstly, we consider an aerial RIS-assisted HAP-terrestrial communication Here, we have considered a HAP serving as an aerial BS, serving ground users within a predefined area. Here, multiple UAVs equipped with RISs assist the links between the HAP and the ground users, when different number of RIS elements, M, is employed. The provided plots show the CDF for SNR and the achievable rate for the users. Additionally, various 3D positions for the UAVs and different heights for the HAP (17km-20km) have been taken into account. Fig. \ref{sim_ntn} demonstrates that RIS-supported scenarios result in a shift towards higher SNR and achievable rate values,
indicating improved communication performance and reliability compared to non-RIS scenarios. The broader spread and higher median values of SNR and achievable rates in RIS-assisted scenarios reflect the robustness and efficiency of RIS in diverse communication
environments. Moreover, increasing the number of RIS elements empowers the RIS to adaptively and intelligently optimize communication links, resulting in improved signal coverage, higher data rates, and enhanced overall system performance. These insights
contribute to the development of efficient and reliable communication solutions for various applications, including aerial communication, IoT connectivity, and beyond.
\begin{figure}[t!]
\centerline{\includegraphics[width=\columnwidth]{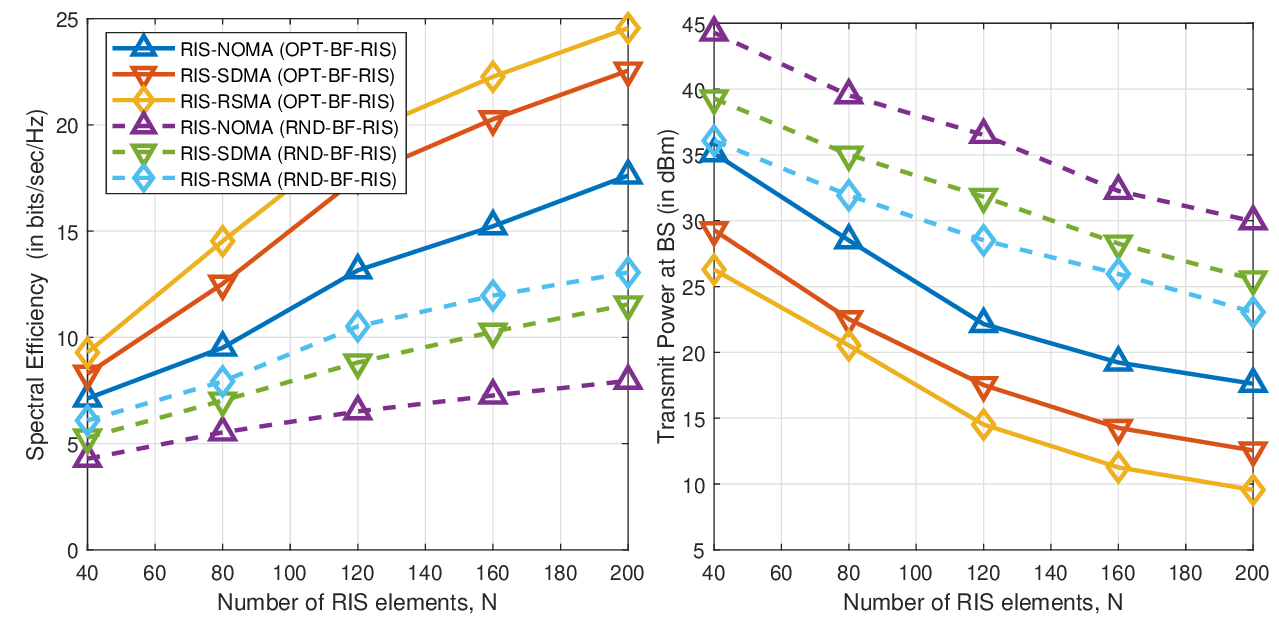}}
	\caption{Performance Evaluation of RIS-aided RSMA  when the minimum QoS rate requirement $=1$bps/Hz, number of users $=6$, number of BS antennas $=4$, maximum transmit power budget $=40$dBm. }
	\label{sim_v2x}
\end{figure}
 \subsection{Interference Managment for V2X using RIS aided Rate-splitting Multiple Access Schemes}
Undoubtedly, interference management is pivotal for ultra-dense networks, where multiple users simultaneously communicate over shared spectral resources, and this becomes particularly crucial in scenarios like V2X, cellular networks, smart cities, etc. Recent studies reveal that the interplay of RIS and rate-splitting multiple access (RSMA) schemes holds promise for effectively mitigating interference. 
As \cite{katwe2022rate}, we examine the performance analysis of the RIS-aided RSMA scheme when compared to other schemes RIS-aided non-orthogonal multiple access scheme  (NOMA) and RIS-aided space division multiple access (SDMA) or conventional multi-user linear precoding for MIMO systems. The results in Fig. \ref{sim_v2x} demonstrate that the RIS-aided  RSMA scheme outperforms both the schemes w.r.t. spectral efficiency as well as transmit power requirement owing to RSMA's better power control than SDMA and NOMA. Also, the performance of the optimal beamforming design (OPT-BF-RIS) surpasses that of random beamforming at the RIS (RND-BF-RIS), and further enhancements are observed with an increase in the number of RIS elements.

\begin{figure}[t!]
\centerline{\includegraphics[width=\columnwidth]{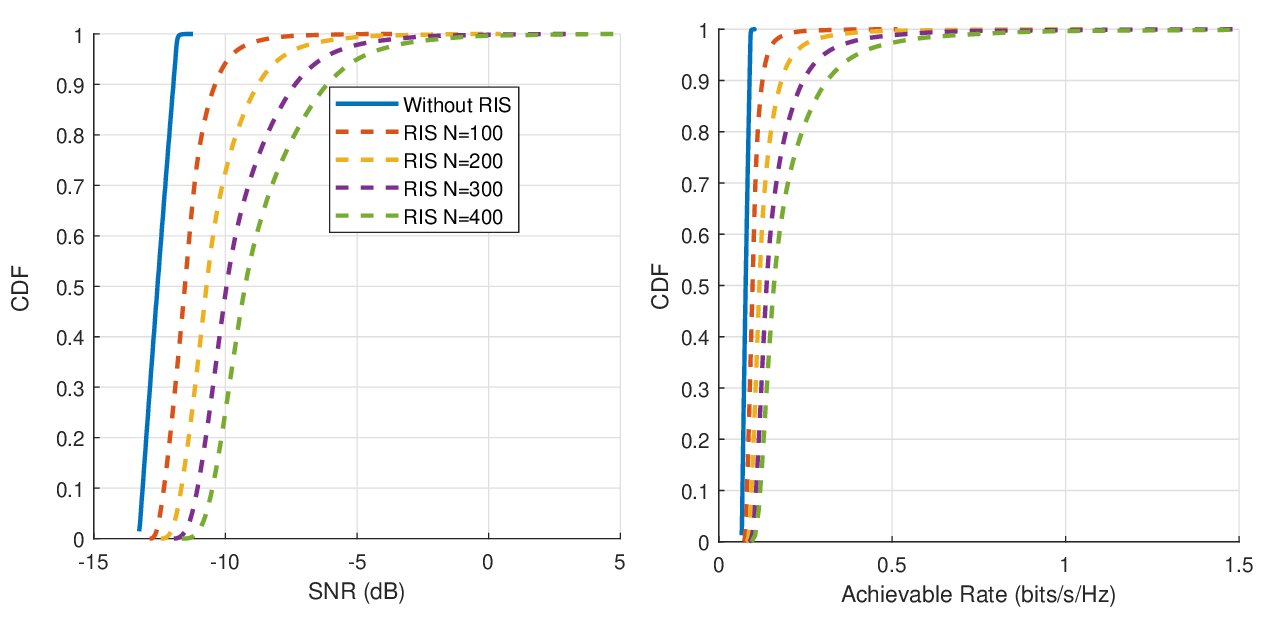}}
	\caption{Performance Evaluation of Aerial RIS-aided HAP to terrestrial communication when  1000 users uniformly distributed across a 10 km$^2$ area and with different heights for the HAP (17km-20km).}
	\label{sim_ntn}
\end{figure}

\section{RIS for 6G and Beyond: Standardization Trends }
\label{sec:3}
Both industry and academia have made significant strides in theoretical development and field testing of IS for 6G , with regional standards developing bodies (SDOs) now actively engaged in standardization efforts.  
\subsection{3GPP and Others: Standardization Trends and Opportunities }
At the October 2020 summit hosted by the International Telecommunication Union (ITU) on radio communication sector, the importance of RIS to the physical layer of future networks was underscored. 
 At the 55th meeting of the Technical Committee 5 — Working Group 6 (TC5 WG6), China Communications Standards Association (CCSA), the proposal to initiate a study item (SI) focusing on RIS garnered approval \cite{liu2022a}. 
Additionally, during the Federal Communications Commission (FCC) Technological Advisory Council (TAC) Meeting on August 17, 2023, it was proposed that the FCC should task a future TAC Working Group to investigate Reconfigurable Intelligent Surfaces (RIS) technologies and assess their potential impact on spectrum allocation and challenges, particularly in the cmWave and mmWave frequency bands. This includes exploring backscatter technology for low-power devices and RIS for high-power commercial mobile systems.

\begin{figure}[t!]
\centerline{\includegraphics[width=0.8\columnwidth]{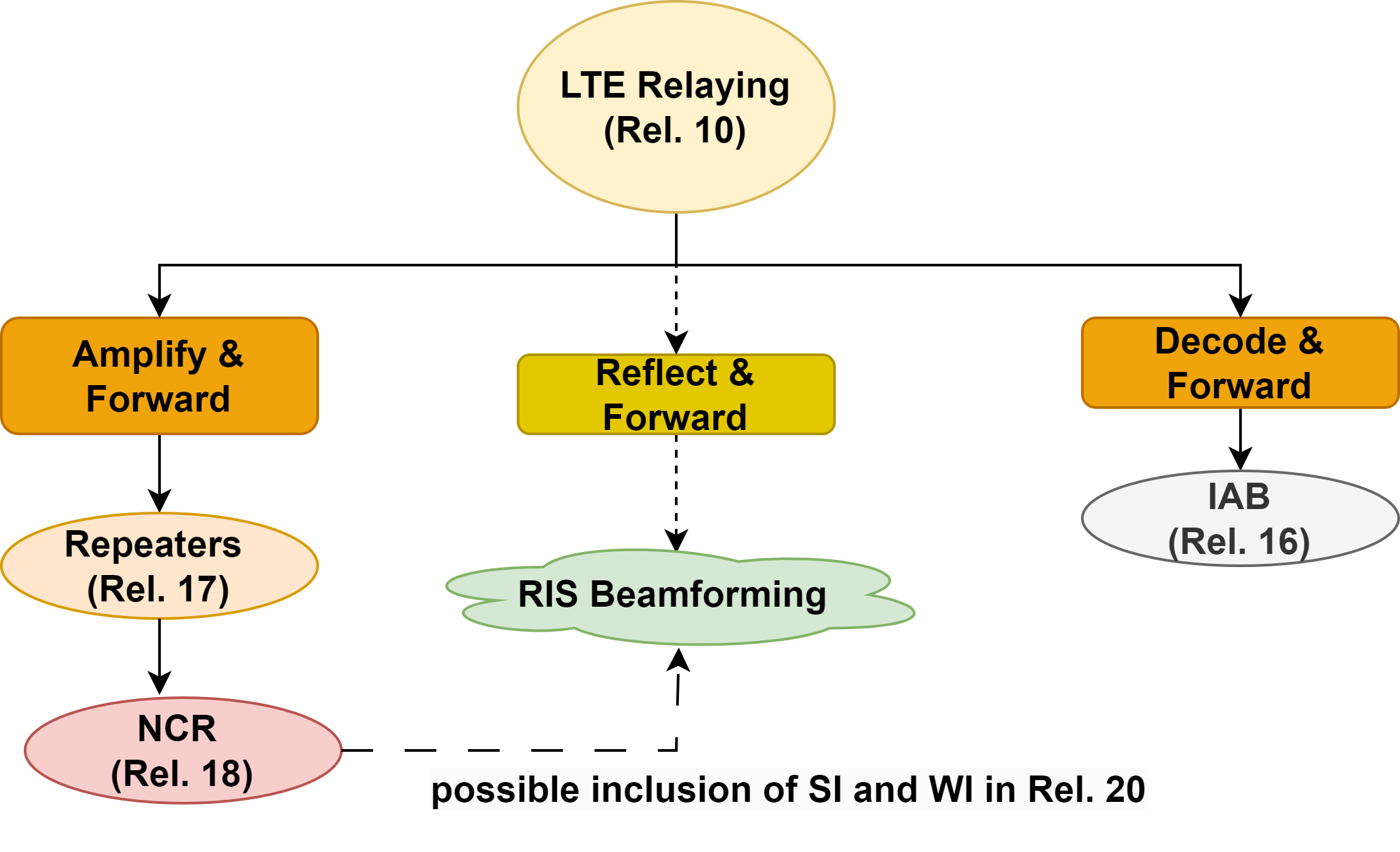}}
	\caption{A possible consideration  of 3GPP  RIS standardization (as proposed by NTT Docomo in Rel. 19 proposal)}
	\label{fig_3gpp}
\end{figure}

Currently, the focus of 3GPP's efforts is on completing Rel. 18 for 5G Advanced systems, with the possibility of including further improvements in Rel. 19, but there are no established plans for 6G and the integration of IS into it at this stage. 
Relaying, which bears close resemblance to RIS, was initially introduced in LTE relaying (Rel. 10), encompassing two main approaches: amplify and forward (AF) and decode and forward (DF). AF relaying, involving repeaters, was explored in Release 17, while DF relaying, including Integrated Access and Backhaul (IAB), was examined in Release 16.  An investigation was conducted as part of 3GPP Release 18 (concluded in August 2022), to assess both the opportunities and obstacles associated with network-controlled repeaters (NCRs).
Building upon the concepts of NCR, another potential relaying option, known as reflect and forward, which forms the foundation of RIS technology, could be introduced in Release 20 as an alternative version of relaying, as illustrated in Fig. \ref{fig_3gpp}.  
For this to materialize, formal approval for the introduction of RIS into standardization is necessary in Rel. 20, requiring the establishment of a study item (SI) and work item (WI).
For instance, RAN1 in Rel. 19 has agreed on two SI, i.e., study on channel modelling for ISAC in NR (RP-234069) and study of channel modelling enhancements for 7-24 Ghz  (RP-234018). The results of this SIs can be useful for the consideration of RIS as SI in Rel.20, paving the way for its continued evolution until 2030 and its potential role as a foundational technology for 7G. 


%

 
\subsection{Industry Specifications and Trends}

In June 2021, a significant milestone was achieved in integrating IS technology into future communication standards with the official approval of a ETSI Industry Specification Group (ISG) dedicated to RISs. The ISG-RIS commenced its activities in September 2021, 
Following the establishment of the ETSI RIS ISG, several key documents have been released to address various aspects of Reconfigurable Intelligent Surfaces (RIS) technology. ETSI GR RIS 003 V1.1.1 (August 2023) focuses on the technological challenges associated with deploying RIS entities, including their potential impact on network architecture, framework, and required interfaces. It also discusses potential specification impacts to support RIS as a new network entity. ETSI GR RIS 001 V1.1.1 (April 2023) outlines use cases, deployment scenarios, and requirements related to Reconfigurable Intelligent Surfaces (RIS), while ETSI GR RIS 002 V1.1.1 (June 2023) covers communication models, channel models, channel estimation techniques, and evaluation methodology specific to RIS technology. These documents represent significant contributions to the standardization efforts surrounding RIS technology and provide valuable insights into its potential applications, deployment considerations, and technical requirements.
These outputs will undergo subsequent evaluation by standards organizations such as 3GPP for future 6G releases and ITU-R for IMT-2030 deliverables, including capabilities and evaluation methodology.

\section{Open  Challenges and their Considerations}
\label{sec:5}
Despite the vast potential of IS, several challenges must be addressed to realize its full impact on future wireless networks. The major challenges are discussed below:
\begin{itemize}
\item \textbf{Standardization}: While ISs are acknowledged as a crucial component for the physical layer of 6G networks, the likelihood of establishing them as a focus group in 3GPP or a WI within the ITU remains uncertain.  At present, the 3GPP is dedicated to completing Rel. 18 of the 5G standard and preparing for future developments in Rel. 19, with no formal discussions or plans regarding the introduction of IS for 6G.
As 6G evolves from concept to reality, industry stakeholders, standards organizations, and regulatory bodies are collaborating to define the specifications, protocols, and interoperability standards governing IS deployment. 
\item \textbf{Channel estimation}:   IS-aided systems presents heightened complexity for CSI when compared to conventional active devices, as the passive reflective elements lack active transmission/reception capability, necessitating channel estimation solely at base stations or user terminals. However, leveraging distributed machine learning optimization techniques and AI modeling under partial/no CSI can assist in addressing these challenges by enhancing accuracy and efficiency in channel estimation processes.
\item \textbf{Wideband IS modeling} ensuring widespread deployment and wideband operation enables exceptional performance, combining communication and sensing capabilities that intuitively map three-dimensional connectivity across ground, air, and satellite networks is vital. However, wideband IS poses a significant challenge, particularly in designing and adapting them to operate across wide frequency ranges.  
\item \textbf{IS Deployment Strategies}: To determine the optimal IS placement, various crucial factors need consideration, including the distribution of users, the intended services within the targeted region, and the interference mitigation with existing networks.  
Another significant challenge in deploying RIS involves establishing viable business models, particularly in outdoor settings, such as determining funding sources for deployment and identifying entities responsible for their management and control. A well-planned deployment of IS can significantly minimize unintentional interference among operators and reduce the time required for CSI estimation, especially when employing IS with sensing capabilities. Therefore, 
strategies need to be developed to mitigate or eliminate potential interference caused by IS reflected signals overlapping with other base stations in the vicinity.
\item \textbf{Interoperability and Regulatory Requirement}: As pointed out in the report ETSI GR RIS 001 V1.1.1 (2023-04),  RIS must be able to work seamlessly with various network operators and users, whether it's implemented by the operators themselves or by third-party entities or individuals. Ensuring interoperability is vital, so RIS needs to meet essential capability standards to facilitate this. Additionally, compliance with regional or industry standards is essential and requires clear specification of RIS requirements accordingly.
\item \textbf{Others}: Additionally, ensuring privacy, security, and resilience in IS-enabled networks remains a critical area of research and development.  The adoption of higher frequency spectrum presents significant challenges for signal propagation, characterized by high penetration losses and limited scattering, complicating the design of efficient multi-antenna transceivers for beamforming and optimization.

\end{itemize}


\section{Conclusions}
\label{sec:7}
This paper has provided an in-depth exploration of the integration of IS into the fabric of 6G and beyond networks. By offering a tutorial-style overview of recent advancements in intelligent meta-surfaces and their potential usage scenarios, we have highlighted the transformative potential of IS technology in shaping the future of wireless communication. However, our discussion also underscored the myriad challenges associated with designing, implementing, and standardizing various types of intelligent surfaces. Despite these challenges, the prospects for IS technology remain promising, with opportunities for enhancing network performance, efficiency, and connectivity. Moving forward, continued research, collaboration, and innovation will be essential for overcoming these challenges and realizing the full potential of IS in future communication systems beyond 6G.
\bibliographystyle{IEEEtran}

\bibliography{IEEEabrv,BibRef}


\end{document}

%% file: Acronym.tex
\acrodef{CSI}{channel state information}
\acrodef{DMA}{dynamic metasurfaces antenna}
\acrodef{EM}{electromagnetic}
\acrodef{IoT}{Internet of things}
\acrodef{IS}{intelligent surface}
\acrodef{NCR}{network control repeater}
\acrodef{NLOS}{non-line-of-sight}
\acrodef{MIMO}{multiple-input multiple-output}
\acrodef{OFDM}{orthogonal frequency division multiplexing}
\acrodef{RIS}{reconfigurable intelligent surface}
\acrodef{SCM}{self-conjugating metasurface}